# Temperature-dependent excitonic photoluminescence Excited by Two-Photon Absorption in Perovskite $CsPbBr_3$ Quantum Dots

**Ke Wei,[1,4,5] Zhongjie Xu,[1,2,4,5] Runze Chen,[2] Chao Shen,[1,4,5] Xiangai Cheng,[1,2,4,5] Tian Jiang[1,2,3,4,5*]**

[1]*College of Opto-Electronic Science and Engineering, National University of Defense Technology, Changsha, 410073, China.*

[2]*State Key Laboratory of High Performance Computing, National University of Defense Technology, Changsha, 410073, China.*

[3]*State Key Laboratory of Low-Dimensional Quantum Physics, Department of Physics, Tsinghua University, Beijing, 10000, China.*

[4]*Hunan Provincial Key Laboratory of High Energy Laser Technology, Changsha, 410073, China.*

[5]*Hunan Provincial Collaborative Innovation Center of High Power Fiber Laser, Changsha, 410073, China.*

[*]*Corresponding author: tjiang@nudt.edu.cn*

**Abstract:** Recently lead halide nanocrystals (quantum dots) have been reported with potential for photovoltaic and optoelectronic applications due to their excellent luminescent properties. Herein excitonic photoluminescence (PL) excited by two-photon absorption in perovskite $CsPbBr_3$ quantum dots (QDs) have been studied across a broad temperature range from 80K to 380K. Two-photon absorption has been investigated with absorption coefficient up to 0.085 cm/GW at room temperature. Moreover, the photoluminescence excited by two-photon absorption shows a linear blue-shift (0.25meV/K) below temperature of ~220K and turned steady with fluctuation below 1nm (4.4meV) for higher temperature up to 380K. These phenomena are distinctly different from general red-shift of semiconductor and can be explained by the competition between lattice expansion and electron−phonon coupling. Our results reveal the strong nonlinear absorption and temperature-independent chromaticity in a large temperature range from 220K to 380K in the $CsPbX_3$ QDs, which will offer new opportunities in nonlinear photonics, light-harvesting and light-emitting devices.

## 1. Introduction

Semiconductor quantum dots (QDs) have been studied intensively as future optoelectronic materials due to their unique properties [1-3]. Due to the quantum confined effect, QDs generally show stable excitons or hole-electron pair at room temperature, large absorption cross sections and strong oscillator strength [4], which will contribute to size tunable band gap [5], narrow emission line width [6], high PL quantum yields (Qys) [7] and multiple exciton generation [8]. Previous researchs were mostly concentrated on metal chalcogenide nanocrystals. Recently lead halide QDs ($MPbX_3$. M =$CH_3NH_3$, Cs; X = Cl, Br, I) have been reported as a new family of nanomaterial. Compared to metal chalcogenide QDs, $MPbX_3$ QDs presented higher PL quantum yields up to 90% [9] without any further surface treatments and temperature-independent chromaticity from 25~100℃ [10], making them a promising alternative to prototypical CdSe QDs for light-harvesting and -emitting applications [11].

In this work, excitonic photoluminescence excited by two-photon absorption is studied at various temperatures from 80K to 380K. When excited by an 800nm femtosecond laser beam, the 11.4nm-sized $CsPbBr_3$ QDs show a photoluminescence with wavelength centered at 513nm and intensity quadratic dependence on excitation density, indicating a two photon absorption. From Z-scan measurement, two photon absorption coefficient up to 0.085 cm/GW is found at room temperature. For more intrinsic characteristic of the generated exciton, photoluminescence under various temperature are carried out. The emission peaks show an unusual linear blue-shift with increasing temperature below ~220K, which can be attribute to the negligible electron−phonon coupling.

Namely, thermal expansion dominates the band gap behaviors. For higher temperature up to 380K, photoluminescence rises to a stable value with fluctuation less than 1nm, indicating a balance between thermal expansion and electron–phonon coupling. To confirm this viewpoint, time-resolved photoluminescence measurement is carried out and show that a maximum exciton lifetime at room temperature (300K), indicating the balance of exciton thermal motion and nonradiative recombination, roughly agreeable with the viewpoint above.

## 1. Result and discussion

### *2.1. Sample preparation and characteristic*

$CsPbBr_3$ QDs are synthesis as previous reports [9]. The inset of Fig. 1(a) shows transmission electron microscopy (TEM) image of the solution QDs (solvent of toluene). The average edge length of $CsPbBr_3$ cube is 11.4nm, comparable with the Bohr diameter (~7nm) of the excitons in bulk counterparts [9]. Quantum confined effect leads to blue-shift of band gap and discrete electric and hole states. Optical transitions between these states result in discrete peaks in linear absorption spectrum, as seen in Fig. 1(a). photoluminescence excited by both 400nm and 800nm femtosecond laser show that the band gap is around 513nm (2.42eV), consistent with previous report [9]. The PL excited by 800nm laser indicates that the $CsPbBr_3$ QDs are indeed multiphoton active [4]. The PL decay is also measured by time correlated single photon counting (TCSPC) with 200ps FWHM of instrument response function (IRF). As shown in Fig. 1(b), almost equal lifetime (~4ns) of excitons generated by 400nm and 800nm laser beam indicates that the recombination mechanisms are the same for two excited wavelength. The deviation after 5ns is attribute to the response of the instrument since the excited frequency and thus measurement range are different for the two excitations.

*2.2. Two photon absorption at room tempertature*

For more insight into the nonlinear absorption of $CsPbBr_3$ QDs, excitation density dependent PL spectrum is carried out, as shown in Fig. 2(a). The sample is held in a spinning sapphire curvette to prevent the nonuniform distribution of QDs and optical tweezers effect. The integrated PL intensity is extracted in Fig. 2(b), the almost quadratic dependence of the PL intensity on the excitation density clearly shows the two-photon absorption in the QDs. Two-photon absorption is a third-order nonlinear process, which simultaneously absorbs two photons through a virtual state to excite one carrier, as shown in the inset of Fig. 2(b).

To quantitatively describe the two-photon absorption, Z-scan measurement was carried out. Details of the set up can be found in Ref [12]. As shown in Fig. 3, the response of the toluene solvent in the maximum intensity is negligible. So the signal most comes from $CsPbBr_3$ QDs. The reverse saturable absorption (RSA) signal in all excitation intensities higher than 40GW/cm$^2$ clearly show nonlinear absorption in the QDs. By fitting the experimental data with Z-scan theory [12], a two-photon absorption coefficient up to ~0.085 cm/GW is obtained, comparable to recent report of $CsPbBr_3$ nanocrystals [4].

*2.3. Temperature-dependent excitonic photoluminescence*

For more essential characteristics of the excitons generated by two-photon absorption, the solution $CsPbBr_3$ QDs are dripped on an ultrathin glass substrate and the stable-state PL spectrum is measured under different temperature from 80K to 380K, as shown in Fig. 4. To ensure single exciton process in carrier relaxation, the excitation intensity (800nm) is adjusted to as low as 0.08mJ/cm$^2$, at which no saturation is found in the PL intensity as shown in Fig. 2(b). Importantly, no significant difference of the PL spectrums is found during the cooling and heating process, providing a credible and reversible result.

At all temperatures as shown in Fig. 4(a), the PL spectrums always show single narrow peaks which are attribute to the emissions of lowest state excitons, indicating a negligible trap states. These negligible trap states may contribute to high conversion efficiency in photoelectric device. PL intensities are normalized to clarify the shift of PL peak, which is summarized in Fig. 4(b). The peak wavelength of the emission shows a linear blue-shift with increasing temperature below 220K. From linear fitting the temperature coefficient is determined as ~0.25meV/K. The blue-shift is different from typically observed of red-shift for semiconductors and can be also observed in recently report of $CsPbBr_3$ nanowires [13], $CH_3NH_3PbI_{3-x}Cl_x$ films [14] and other semiconductors like closely related cesium metal halide [15], PbS (Se/Te) and CuCl (Br/I) QDs [16] and Pb-doped CsBr crystals [17].

It's well known that thermal expansion and electron-phonon interaction will contribute to the modification of semiconductor band gap. Typically electron-phonon interaction dominates the band gap behavior and often gives a red-shift to the band gap, which can be well described by traditional empirical Varshni models. However in some materials, as Yu reported [15], there are probably two phonon modes which have opposite contributions to band gap and thus they counteract each other. So the thermal expansion dominates the band gap behavior and they may attribute a linear blue-shift to the band gap. Domination factor below temperature of 220K is thermal expansion, which often gives a linear blue-shift to the band gap with increasing temperature. Similarly behavior is hypothesized in $CsPbBr_3$ QDs for temperature under 220K.

The emission peak increases to a maximum at temperature of 220~300K. Then it shows a very shallow trough with fluctuation less than 1nm (4.4meV) around 330K. The relatively steady band gap is probably attribute to the electron-phonon interactions, which can't no longer be ignored at high temperature. Thus competition between the two factors will result in complex behaviors of band gap. Whatever, the stable emission peaks of $CsPbBr_3$ QDs was agree with recent study [10], at which PL peak positions do not alter from 25 to 100℃.The temperature-independent chromaticity is particularly advantageous to applications such as light-emitting diode, which often heats up during prolonged operation.

Fig. 4(b) also shows the evolution of PL line width (FWHM), which presents a monotonous broadening with increasing temperature and can be commonly described using the following model [18]:

$$\Gamma(T) = \Gamma_0 + \Gamma_{\text{photon}}(e^{\hbar\omega_{\text{LO}}/k_B T} - 1)^{-1} + \Gamma_{\text{imp}} e^{-E_b/k_B T} \quad (1)$$

Where the first term $\Gamma_0$ is the inhomogeneous broadening contribution. The second term describes the coupled of longitudinal optical phonon to the electronic transition. The $\Gamma_{\text{photon}}$ and $\hbar\omega_{\text{LO}}$ represent the coupling strength and longitudinal optical phonon energy, respectively. The third term is attribute to ionized impurity scattering which can be neglected in $CsPbBr_3$ QDs due to the negligible trap states as described above. From fitting process, estimated values of the parameters are obtained as follows: $\Gamma_0$ , $\Gamma_{\text{photon}}$ = 32(2) , 130(20) meV, and $\hbar\omega_{\text{LO}}$ = 30(3) meV.

Finally, to reconfirm the above model which has a contribution to the blue-shift of PL spectrum, temperature dependent PL lifetime is carried out, as shown in Fig. 5. The lifetime increase continuously below room temperature and goes to a maximum(2.4ns) in ~300K, agreeing well with the behavior of the emission peak, which also shows a peak at 220~300K (Fig. 4(b)). Because the QDs are frozen on a glass substrate, the lack of Brownian motion results in a smaller PL lifetime (2.4ns) at 300K compared to solution QDs (~4ns). For temperature below 300K, due to small electron-phonon interaction, the nonradiative recombination is relatively small and the radiative recombination of single exciton state dominates the carrier decay. Thus with increasing temperature, thermal motion becomes more active and thus prevents the recombination of excitons, resulting in an increased PL lifetime, as shown in the below column of Fig. 5. While for temperature higher than 300K, as shown in above column of Fig. 5, the increasing electron-phonon interaction will drastically increase the nonradiative recombination, which will reduce PL lifetime. This behavior is commonly known as the thermal quenching. On the other hand, The transition from exponential to non-exponential decay at 300K (above column of Fig. 5) also indicates an increasing nonradiative process in $CsPbBr_3$ QDs.

## 2. Conclusion

In conclusion, excitonic photoluminescence excited by two-photon absorption in perovskite $CsPbBr_3$ QDs is studied at a wide temperature range from 80K to 380K by steady and time-resolve PL spectroscopy and Z-scan measurement. Two-photon absorption coefficient up to 0.085 cm/GW is found at room temperature. For temperature below 220K, The PL peak shows an uncommon linear blue-shift (0.25meV/K) with increasing temperature, indicating that the thermal expansion dominates the band gap behaviors, while the electron-phonon interaction is negligible. For higher temperature, a stable emission peak in temperature range from 220K to 380K is found and is attribute to the balance between thermal expansion and electron-phonon interaction. The strong nonlinear absorption and the luminescence monochromaticity in a wide temperature range make the $CsPbBr_3$ QDs promising materials in low-cost nonlinear absorber, light-harvesting and light-emitting applications.

**Funding.**

National Natural Science Foundation of China (NSFC) (61340017)；The Scientific Researches Foundation of College of Optoelectronic Science and Engineering, National University of Defense Technology (No. 0100070014007).

**Acknowledgment.**

We would like to thank Doctor L.F.Jiang (Nanjing MKNANO Tec. Co.,Ltd., China) for providing $CsPbBr_3$ QDs and its linear absorption spectrum and TEM image.

**Figures**

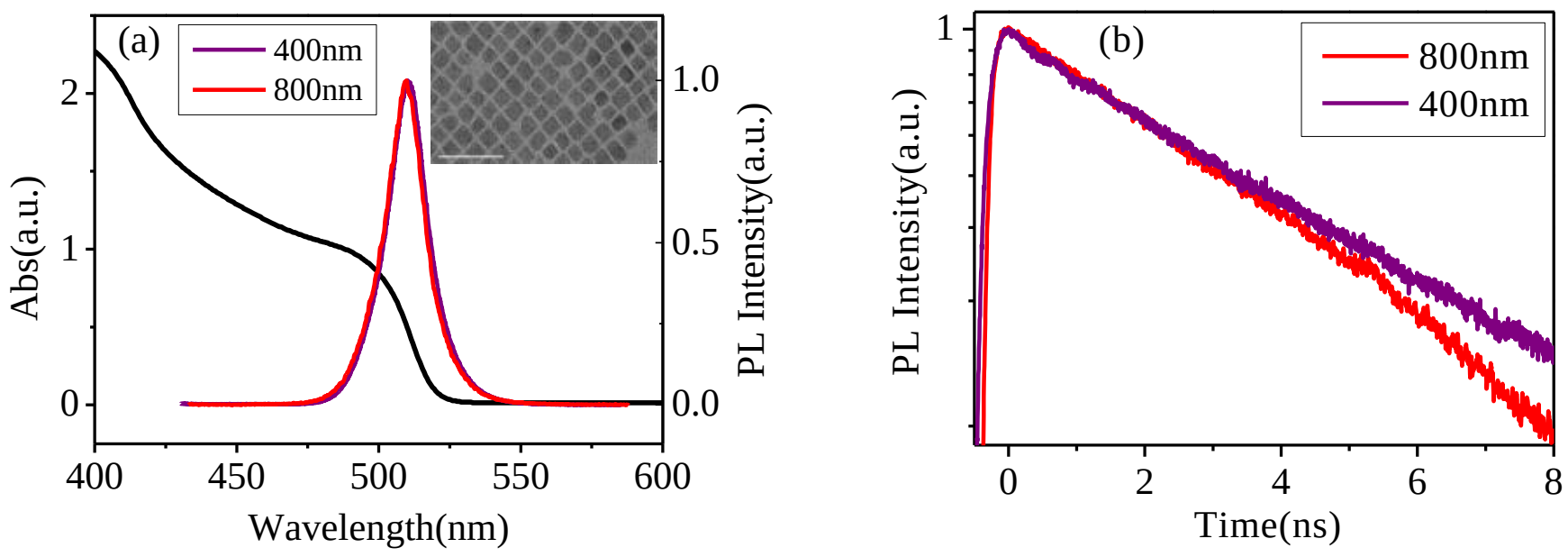

Fig. 1. Characterization of $CsPbBr_3$ QDs. (a) Linear absorption(black) and photoluminescence excited by 400nm (purple) and 800nm (red) laser. Inset shows the TEM image with scale bar of 50nm. (b)PL decay excited by 400nm (purple) and 800nm (red) laser.

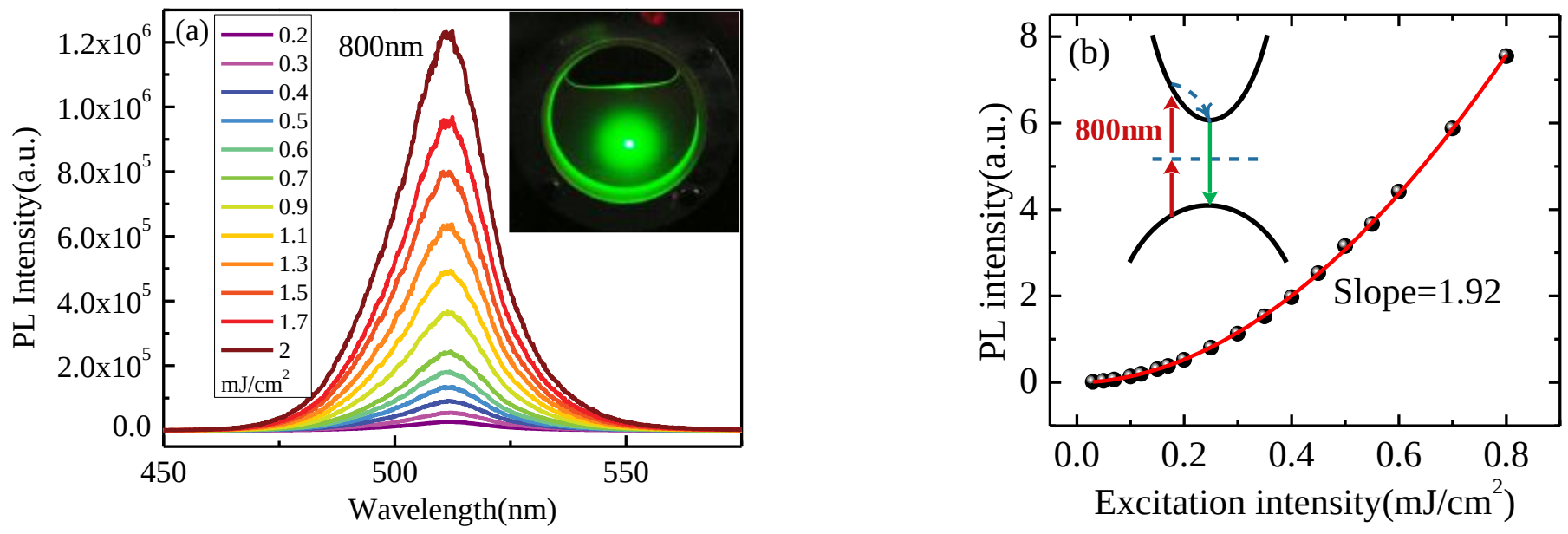

Fig. 2. (a) PL spectrum under different excitation densities at room temperature. The inset shows the photograph of $CsPbBr_3$ QDs solution illuminated by 800nm laser beam. (b) integrated PL intensity as a function of excitation intensity. The inset shows the schematic of two photon absorption.

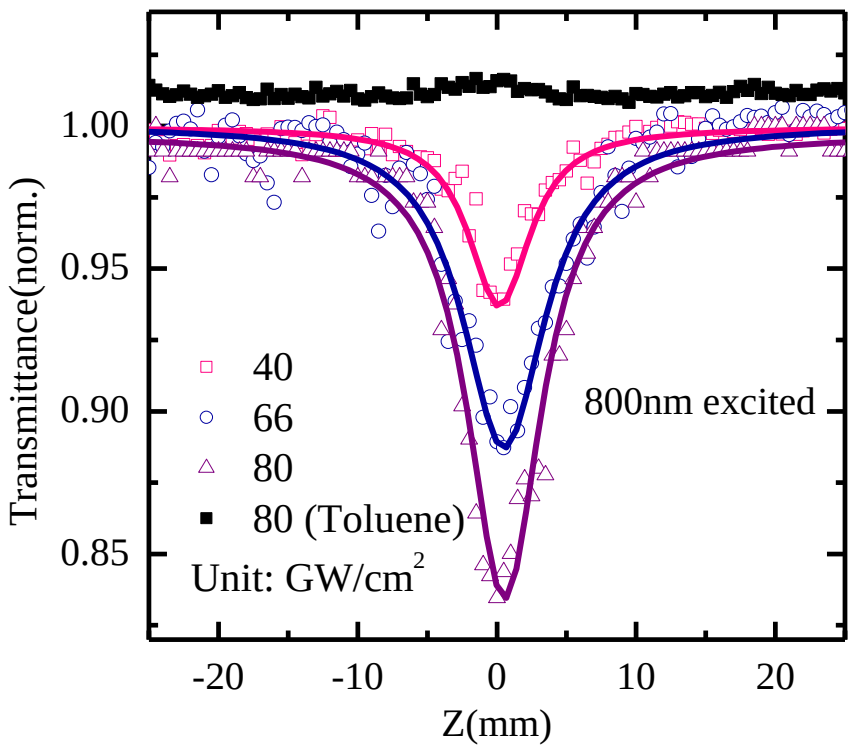

Fig. 3. Open aperture Z-scan measurement for $CsPbBr_3$ QDs and the pure solvent (Toluene) at different excitation intensities (at focus). The scatters are experimental results and the solid lines show the best fits using Z-scan theory.

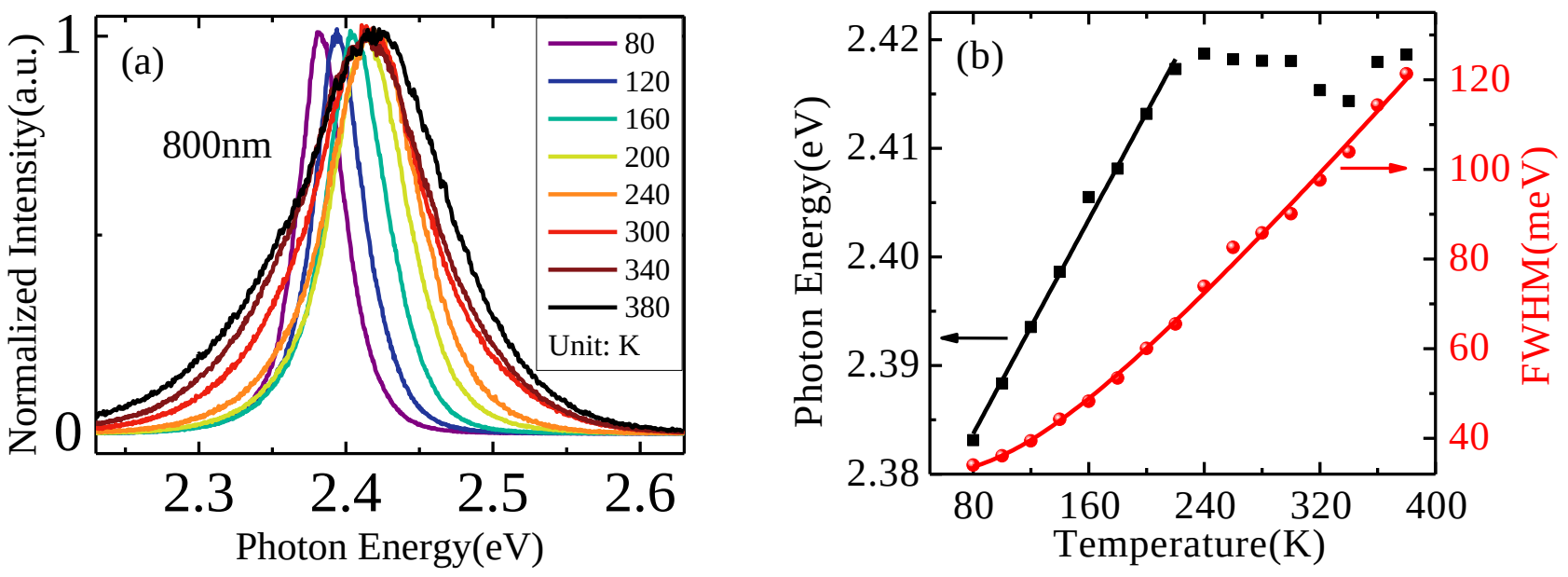

Fig. 4. (a) Temperature dependent PL spectrum excited by 800nm laser beam. (b) The peak (black) and line width (red) of the PL spectrum as a function of excitation intensity. The scatters are experimental results and the solid lines are fitting curves.

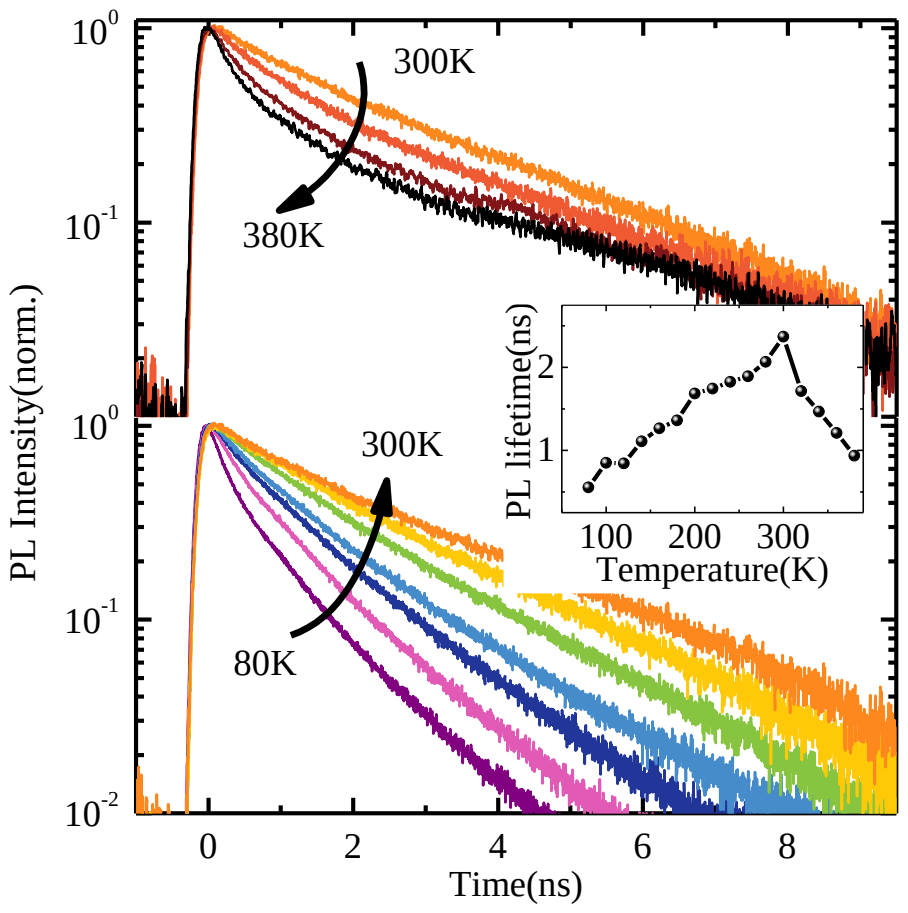

Fig. 5. Temperature dependent of PL decay. The above and below columns show decay curve in temperature ranges of 300 ~380K and 80~380K, respectively. Inset extracts the effective PL lifetime from the two columns.